\documentclass[a4paper,11pt]{article}
\usepackage{pos}
\usepackage{comment}

\usepackage[numbers,sort&compress,unsrt]{}
\title{Differential Higgs production at small-$x$ }
\ShortTitle{Differential Higgs production at small-$x$}

\author*[a, b]{Benedetta Bernardini}

\affiliation[a]{Dipartimento di Fisica, Sapienza Università di Roma\\
 Piazzale Aldo Moro 5, 00185 Roma, Italy}

\affiliation[b]{INFN, Sezione Roma 1\\
Piazzale Aldo Moro 5, 00185 Roma, Italy}

\emailAdd{benedetta.bernardini@uniroma1.it, benedetta.bernardini@infn.roma1.it}
\emailAdd{}

\abstract{
In the context of precision physics, the resummation of large logarithms of the variable \(x = Q^2/s\) is increasingly relevant at high energies. We report here preliminary results on the extension of the differential resummation formalism within the HELL framework to Higgs boson production via gluon fusion \cite{nostro}.
}

\FullConference{
DIS 2026, Bologna (Italy), 4-8 May 2026
}

\begin{document}
\maketitle



\section{Introduction}
The increasing precision of experimental measurements requires theoretical predictions with comparable accuracy. This requires not only higher-order fixed-order calculations, but also an accurate description of parton distribution functions (PDFs) and the resummation of logarithmically enhanced contributions. At high collider energies $\sqrt{s}$, hadron collisions probe smaller values of the partonic momentum fraction $x \sim Q^2/s$. In this regime, the perturbative expansion receives contributions of the form $\alpha_s^n \ln^n(1/x)$, which can spoil the convergence of the fixed-order expansion and therefore require all-order resummation.
Such high-energy logarithms arise both in PDF evolution and in partonic coefficient functions. Hence a consistent small-$x$ prediction requires both contributions to be resummed. 
The theoretical framework for small-$x$ resummation in PDF evolution has been developed over the past three decades \cite{Lipatov:1976zz, Salam_1998, Fadin_1998, PhysRevD.60.114036, Ciafaloni_2007, Ball_1995, Altarelli_2008, Thorne_2000, Ball:2017otu}. 
Instead the resummation of partonic coefficient functions is based on high-energy ($k_t$-) factorization~\cite{Catani-Hautmann, CATANI_CIAFALONI}, in which the incoming partons are treated as off-shell.


For inclusive Higgs production, the numerical impact of small-\(x\) resummation at future colliders is sizeable but is found to be dominated by the resummation of the PDFs \cite{Bonvini_2018}. Differential observables can provide enhanced sensitivity to regions of phase space in which the partonic coefficient function contribution becomes more relevant.
Differential small-$x$ resummation was originally formulated for observables such as rapidity and transverse-momentum distributions~\cite{Muselli:2017fuf,Caola_2011,Forte_2016}. Then a stable numerical implementation was developed within the \texttt{HELL} framework~\cite{Bonvini_2016,Bonvini_2017,Bonvini_2018}, that recently has been applied to heavy-quark pair production~\cite{Silvetti_2023}.
In this work, we extend the differential implementation in \texttt{HELL} to Higgs boson production via gluon fusion. The kinematics of the Born process lead to particularly simple relations between the Higgs momentum and the transverse momenta of the incoming off-shell gluons, while the loop induced nature of the Higgs-gluon coupling requires the quark-mass dependence to be retained. 


\section{Small-$x$ resummation in the \texttt{HELL} language for differential coefficient functions}



We consider the cross section differential in hadronic rapidity $Y$ and transverse momentum $q_t^2$ of the final state. It is written in the usual collinear factorization as
\begin{equation}
\begin{aligned}
      \frac{d\sigma}{dYdq_t^2}\left(\tau, Y, Q^2, q_t^2 \right) &= \tau \sum_{ij}\int_{\tau}^1 \frac{dx}{x}\int dy {\frac{dC_{ij}}{dydq_t^2}\left(x,y, Q^2, q_t^2, \alpha_s, \frac{Q^2}{\mu_F^2} \right)}\\
       & \quad \times {f_i\left(\sqrt{\frac{\tau}{x}}e^{Y-y},\mu_F^2\right)f_j\left(\sqrt{\frac{\tau}{x}}e^{-(Y-y)},\mu_F^2\right)};
       \label{sigma_collinear}
\end{aligned}
\end{equation}
where \( x_{1,2}=\sqrt{\tau/x} \ e^{\pm (Y - y)}
\)
and $\tau = Q^{2} / s$. The sum $i,j$ runs all over the different partons, and the coefficient function $C$ is computed in the center of mass of the initial partons.

In the small-$x$ limit, the cross section can be factorized using the $k_t$-factorization \cite{Catani-Hautmann, CATANI_CIAFALONI}. In this formulation the incoming gluons carry transverse momenta $\boldsymbol{k_{t1}}$ and $\boldsymbol{k_{t2}}$ and the cross section takes the form
\begin{equation}
\begin{aligned}
     \frac{d\sigma}{dYdq_t^2}\left(\tau, Y, Q^2, q_t^2 \right)&=\tau \int_\tau^1 \frac{dz}{z} \int d\eta \int_{0}^{\infty} dk_{t1}^2 \int_{0}^{\infty} dk_{t2}^2 {\frac{d\mathcal{C}}{d\eta dq_t^2}(z,\eta, k_{t1}^2, k_{t2}^2, Q^2, q_t^2)}\\
     & \quad \times {\mathcal{F}_g\left(\sqrt{\frac{\tau}{z}}e^{Y- \eta},k_{t1}^2\right)\mathcal{F}_g\left(\sqrt{\frac{\tau}{z}}e^{-(Y- \eta)},k_{t2}^2\right)}.
     \label{sigma_kt}
\end{aligned}
\end{equation}
The off-shell coefficient function $\mathcal{C}$ is computed with incoming gluons carrying transverse momentum. It is two-gluon-irreducible (2GI), so that the high-energy logarithms are factorized into the unintegrated gluon distributions (UGDs) $\mathcal{F}_g (x, k_t^2)$.
Within the \texttt{HELL} framework \cite{Bonvini_2016, Bonvini_2017, Bonvini_2018}, the UGDs are related to the collinear PDFs  through the universal evolutor $U_{\text{gg}}'$:
\begin{equation}
     \mathcal{F}_g(x, k_{t}^2)= \int_x^1 \frac{dz}{z}U^{'}_{gg}(z, k_t^2, \mu_F^2)f_g\left(\frac{x}{z},\mu_F^2\right);
     \quad  U'_{\text{gg}}(z, k_t^2, \mu_F^2)=\frac{d}{dk_t^2}U_{\text{gg}}(z, k_t^2, \mu_F^2).
    \label{eq:UGD}
\end{equation}
At LL accuracy, the relation between UGD and the collinear gluon PDF is given by eq.~\eqref{eq:UGD}, with the evolutor $U_{gg}$ defined such that the relevant small-$x$ logarithms are resummed to all orders \cite{Bonvini_2016, Bonvini_2017, Bonvini_2018, Silvetti_2023}.
Substituting eq.~\eqref{eq:UGD} into the $k_t$-factorization formula yields the resummed partonic coefficient function
\begin{equation}
    \begin{aligned}
    {\frac{dC_{gg}}{dydq_t^2}\left(x,y, Q^2, q_t^2, \alpha_s, \frac{Q^2}{\mu_F^2} \right)}&=\int_0^{\infty} dk_{t1}^2 \int_0^{\infty} dk_{t2}^2 \int_x^1 \frac{dz}{z}\int d\bar{\eta}\ \frac{d\mathcal{C}}{d\eta dq_t^2}(z,y-\bar{\eta}, k_{t1}^2,k_{t2}^2, Q^2, q_t^2)\\
    &\quad \times {U^{'}_{gg}\left(\sqrt{\frac{x}{z}}e^{\bar{\eta}} ,k_{t1}^2, \mu_F^2\right)} {U^{'}_{gg}\left(\sqrt{\frac{x}{z}}e^{-\bar{\eta}}  ,k_{t2}^2, \mu_F^2\right)} ;
    \label{dC_res}
\end{aligned}
\end{equation}
where \(\bar{\eta} = y - \eta\). This expression provides the basis for the implementation of the differential coefficient function resummation in \texttt{HELL}. The process dependence is entirely contained in the off-shell coefficient function $\mathcal C$, while the evolutors are universal functions.

\section{Higgs production}
The dominant Higgs production mechanism at hadron colliders is gluon fusion, $gg \to H$. Hence, in the present application, the Born-level process with off-shell incoming gluons is a $2\to1$ process. Consequently, the kinematics of the produced Higgs momentum $q^\mu = (q^0, \boldsymbol{q_t}, q^3)$ are completely determined by the momenta  $k_1^\mu, k_2^\mu$ of the two incoming gluons. In particular,
\begin{equation}
    \boldsymbol{q_t} = \boldsymbol{k_{t1}} + \boldsymbol{k_{t2}};
    \qquad q_t^2 = |\boldsymbol{k_{t1}} + \boldsymbol{k_{t2}}|^2;
\end{equation}
while the on-shell condition fixes the invariant mass
\begin{equation}
\begin{split}
  Q^2 &\equiv q^2=m_H^2=(k_1+k_2)^2=x_1x_2s-q_t^2;\\
    \eta&\equiv\frac{1}{2}\log{\frac{q^0+q^3}{q^0-q^3}}-\bar{\eta}=0.
\end{split}
\end{equation}
Here $\eta$ is the relative rapidity between the Higgs and the partonic system in the partonic center-of-mass frame.
These relations considerably simplify the computation of the off-shell coefficient function in eq.~\eqref{dC_res}.

The Higgs kinematics also impose an upper bound on the partonic variable $x$ entering the coefficient function convolution. This variable should be distinguished from the hadronic momentum fractions $x_{1,2}$ appearing in eq.~\eqref{sigma_collinear}.
At fixed Higgs transverse momentum, this bound is \[x < x_{\max} = \frac{1}{1 + \left(\frac{q_t}{m_H}\right)^2}.\]
This simple kinematic constraint provides a useful interpretation of the relative importance of PDF and coefficient-function resummation. For $q_t \ll m_H$, one has $x_{\max} \approx 1$, so the coefficient function convolution samples a broad range of partonic momentum fractions and the dominant small-\(x\) effects are expected to originate from the resummed PDFs. As \(q_t\) increases, \(x_{\max}\) decreases, restricting the convolution to smaller values of \(x\) and thereby increasing the sensitivity to the high-energy behaviour of the off-shell coefficient function.
The rapidity dependence provides an additional handle. Around central rapidities , $Y \approx 0$, the partonic momentum fractions are approximately symmetric, $x_1 \approx x_2$; 
while at forward rapidities, increasingly asymmetric momentum fractions are probed. So in the forward region the small-$x$ logarithmic enhancement is increasingly suppressed by the large-\(x\) behaviour of the PDF.

Another important feature of this process is the quark mass dependence. Fixed-order differential Higgs predictions are known to very high orders in the heavy-top effective theory~\cite{Mistlberger_2018,PhysRevD.99.034004,PhysRevLett.127.072002}. However, the heavy-top limit does not reproduce the correct high-energy behaviour of the off-shell amplitude. The small-$x$ and the heavy-top limits therefore do not commute, and the finite quark-mass dependence must be retained in the computation of the off-shell amplitude before applying the high-energy resummation.

\subsection{Numerical results}
As a first numerical application, we evaluate the resummed differential partonic coefficient function in \texttt{HELL}.
\begin{figure}[t]
    \centering
    \includegraphics[width=0.5\linewidth]{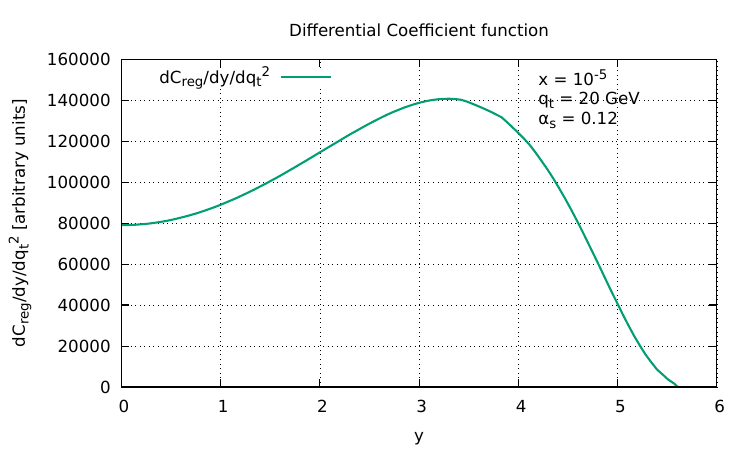}
    \caption{The resummed partonic coefficient function $\frac{dC_{\text{reg}}}{dy dq_t^2}$ as a function of the partonic rapidity $y$ at fixed $q_t = 20\text{ GeV}$, $x = 10^{-5}$, and $\alpha_s = 0.12$ represented in arbitrary units.}
    \label{fig:Cgg}
\end{figure} Figure~\ref{fig:Cgg} shows the result
as a function of the partonic rapidity $y$,  fixing
\(
x=10^{-5},
\
q_t=20~{\rm GeV} \  \rm {and} \ 
\alpha_s=0.12.
\)
It displays a non-trivial $y$ dependence resulting from the convolution of the off-shell Higgs coefficient function with the two evolution operators. The vanishing of the partonic coefficient function at the kinematic boundary provides a useful check of the implementation of eq.~\eqref{dC_res}. 
This calculation provides the process-dependent ingredient required to extend the differential \texttt{HELL} formalism to Higgs production.


Convolving the resummed partonic coefficient functions with the collinear PDFs yields the physical double-differential cross section. 
\begin{figure}[t]
    \centering
    \includegraphics[width=0.5\linewidth]{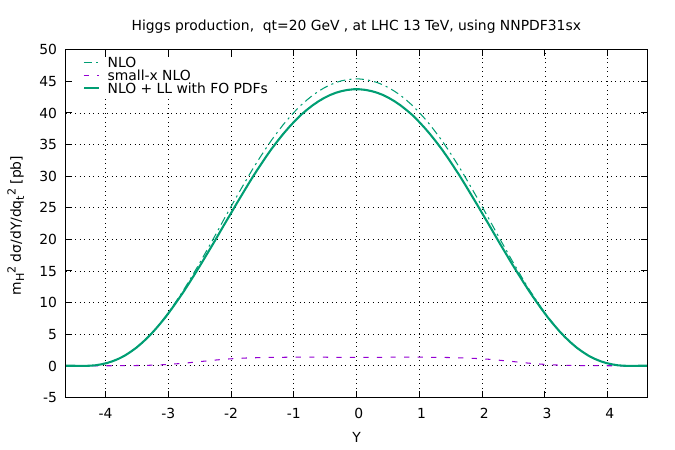}
    \caption{Double-differential Higgs cross section $m_H^2\frac{d\sigma}{dY dq_t^2}$ in pb as a function of rapidity $Y$ at $q_t = 20\text{ GeV}$ at LHC ($\sqrt{s} = 13\text{ TeV}$). Curves compare fixed-order NLO (dash-dotted green), fixed-order small-$x$ expanded NLO (dashed purple), and the fully matched $\text{NLO}+\text{LLx}$ prediction with fixed-order PDFs (solid green).}
    \label{fig:PLOT2}
\end{figure}
Figure~\ref{fig:PLOT2} shows the hadronic distribution as a function of the hadronic rapidity $Y$, fixing $q_t = 20\text{ GeV}$ and $\sqrt{s} = 13\text{ TeV}$, using the \texttt{NNPDF31sx}\ PDF set at fixed-order. We compare the fixed-order NLO prediction, computed using the \texttt{POWHEG-BOX} code \cite{Bagnaschi:2011tu, Alioli:2010xd, Frixione:2007vw, Nason:2004rx}, with its small-\(x\) expansion and with the matched NLO+LL\(x\) result. 
Since the Born contribution to Higgs production is proportional to \(\delta(q_t^2)\), the first non-vanishing contribution at finite transverse momentum appears at NLO. At \(q_t=20\) GeV, the matched NLO+LL\(x\) prediction lies below the fixed NLO result, with the largest relative difference around central rapidity. The effect decreases towards the rapidity endpoints, in the forward and backward rapidity tails, where increasingly asymmetric momentum fractions are probed and the large-\(x\) suppression of one incoming PDF reduces the impact of the small-\(x\) enhancement.

\section{Conclusions}
We have presented preliminary results on the differential small-$x$ resummation formalism implemented in the \texttt{HELL} framework to Higgs boson production via gluon fusion. The calculation combines the high-energy factorization of the partonic coefficient function with the universal small-$x$ evolution encoded in the \texttt{HELL} evolutors. While inclusive Higgs production is largely driven by PDF resummation, differential distributions in rapidity and transverse momentum provide enhanced sensitivity to the partonic coefficient function contribution.
The kinematics of Higgs production provide a particularly simple setting for implementing the off-shell coefficient function, while the loop-induced nature of the process requires the full finite quark-mass dependence to be retained. 
The first numerical results show a sizeable reduction of the fixed-order NLO prediction at $q_t=20$ GeV, with the largest effect around central rapidity. The kinematic upper bound on the partonic variable $x$ provides a simple interpretation of the increased sensitivity to coefficient-function resummation at larger transverse momentum.
Future developments will focus on systematically mapping these kinematic regions at LHC and FCC-hh energies and on quantitatively determining the regions in which coefficient function resummation becomes more important relative to PDF resummation.
\bibliography{biblio}
\end{document}